\documentclass{report}
\usepackage{graphicx}
\usepackage{amstex,amscd,amssymb}
\usepackage{svcon2e}

\begin{document}

\begin{center}
\begin{huge}
 Light quark masses and  condensates   in
  QCD\footnote{Invited 
plenary talk given at the Workshop on Chiral Dynamics 1997, Mainz, 
Germany, Sept. 1-5, 1997}\\
\end{huge}
\thispagestyle{empty}
\vspace{1cm}

\begin{Large}
Jan Stern\\
\end{Large}

\vspace{2cm}

Division de Physique Th\'eorique, Institut de Physique Nucl\'eaire,\\
	Universit\'e Paris-Sud, 91406 Orsay, France\footnote{Unit\'e de 
	Recherche des Universit\'es Paris XI et Paris VI, associ\'ee au CNRS}\\

\vspace{3cm}

\textbf{Abstract}\\
\end{center}
\vspace{.5cm}

	We review some theoretical and phenomenological aspects of the 
	scenario in which the spontaneous breaking of chiral symmetry is not 
	triggered by a formation of a large condensate $<\bar qq>$. Emphasis 
	is put on the resulting pattern of light quark masses, on the 
	constraints arising from QCD sum rules and on forthcoming 
	experimental tests.

\vfill
IPNO/TH 97-30

\pagenumbering{arabic} 
\chapter{Light quark masses and condensates in QCD}
%\footnote{Invited 
%plenary talk given at the Workshop on Chiral Dynamics 1997, Mainz, 
%Germany, Sept. 1-5, 1997}}
\chapterauthors{Jan Stern}
\setcounter{page}{1}
\begin{abstract}
	We review some theoretical and phenomenological aspects of the 
	scenario in which the spontaneous breaking of chiral symmetry is not 
	triggered by a formation of a large condensate $<\bar qq>$. Emphasis 
	is put on the resulting pattern of light quark masses, on the 
	constraints arising from QCD sum rules and on forthcoming 
	experimental tests.
\end{abstract}
\section{Introduction}
In the presence of $N_{f}$ massless flavours, the QCD Lagrangian 
exhibits the chiral symmetry
$SU_{L}(N_{f})\times SU_{R}(N_{f})\times U_{V}(1)$.
Two theoretical facts can be inferred from first principles.
\newtheorem{montheo}{Theorem}
\begin{montheo}
If $N_{c}\geq 3,\ N_{f}\geq 3$, and provided quarks are confined 
(no coloured physical states), the chiral symmetry 
$SU_{L}(N_{f})\times SU_{R}(N_{f})\times U_{V}(1)$ is necessarily 
broken down to its diagonal subgroup $U_{V}(N_{f})$ generated by 
vector currents.
\end{montheo}
This theorem follows from the analysis of constraints imposed by 
anomalous Ward identities on the spectrum of massless physical states 
\cite{t'H}, \cite{Fri}, \cite{Co} and from some non-trivial properties 
of vector-like gauge 
theories, such as the so called "persistent mass condition" \cite{PW}:
a bound state can be massless only if all its constituents are 
massless \cite{VW}, \cite{Vafa84a}. The proof of Theorem 1 gives a hint about the 
actual content of the statement of spontaneous breaking of chiral 
symmetry $(SB\chi S)$: the statement merely concerns the existence of 
massless Goldstone boson states (pions) coupled to conserved 
axial currents:
\begin{equation}
	<0\vert A^{i}_{\mu}\vert\pi^j\vec p>=i\delta^{ij}F_{0}p_{\mu}.
	\label{Api}
\end{equation}
One proves that in QCD with $N_{f}\geq 3$ and no coloured states, it 
would be impossible to satisfy all anomalous Ward identities if 
$F_{0}$ would have to vanish. This brings us to the second 
theoretical fact.
\begin{montheo}
A necessary and sufficient criterion of $SB\chi S$ is a non-zero 
value of the left-right correlation function
\begin{equation}
	\lim_{m\to 0}i\int d^{4}x<\Omega\vert T L^{i}_{\mu}(x)
	R^j_{\nu}(0)\vert\Omega>=-
	\frac{1}{4}\eta_{\mu\nu}\delta^{ij}F^{2}_{0},
	\label{limi}
\end{equation}
where $L_{\mu}=\frac{1}{2}(V_{\mu}-A_{\mu}),\ 
R_{\mu}=\frac{1}{2}(V_{\mu}+A_{\mu})$
are Noether currents generating the left and right chiral rotations 
respectively.
\end{montheo}
The correlator \ref{limi} encompasses both essential features of $SB\chi S$.
{\bf i)} The asymmetry of the vacuum: if the vacuum would be 
symmetric, $F^2_{0}$ should vanish. {\bf ii)} The existence of 
massless Goldstone bosons: the limit \ref{limi} is nonvanishing if 
and only if the correlator  \ref{limi} contains a massless pion-pole.
$F^2_{0}$ is an {\bf order parameter}, whose non-zero value is 
necessary and sufficient for $SB\chi S$ and Goldstone bosons to occur.
There are, of course, many other order parameters, such as local quark 
condensates
$<\bar qq>,\ <\bar q\sigma_{\mu\nu}F^{\mu\nu}q>\ldots$
A non-zero value of each of them by itself implies $SB\chi S$, but 
the converse is not true. $SB\chi S$ can take place (i.e. $F_{0}\neq 
0$), even if {\bf some} of these condensates vanishes. In particular, 
there is no proof available as in the case 
of $F^2_{0}$, showing that $<\bar qq>\neq 0$ is a necessary 
consequence of $SB\chi S$. The $\bar qq$ condensate plays an 
analogous role as the spontaneous magnetization $<\vec M>$ of spin 
systems with broken rotation symmetry. Although there is no symmetry 
reason for the latter to vanish, its actual value depends on the 
nature of the magnetic order in the ground state:
$<\vec M>\neq 0$ for a ferromagnet, whereas  $<\vec M>=0$ for an 
antiferromagnet.
 
In the next section, I shall briefly illustrate how $SB\chi S$ without 
$<\bar qq>$ condensation could naturally arise in QCD. The existence 
of such a theoretical possibility means that one should remain open-minded 
and precautious concerning the value of $<\bar qq>$, especially, since 
the latter is not yet under an experimental control. I shall mainly 
review the pattern of light quark masses as they would look like if 
the condensate was considerably smaller than usually believed, adding 
some comments on the constraints imposed on quark masses and 
condensates by QCD-sum rules. Finally, I shall briefly mention few 
forthcoming experimental tests. Some of them are discussed in details 
in other contributions to this Workshop \cite{Eck}, \cite{Meiss}, 
\cite{sain}, \cite {Knecht}, \cite{Lee97}, \cite{Lowe97}, 
\cite{Schacher97}.
\section{QCD vacuum as a disordered system}
Upon  evaluating non-perturbative quantities such as the correlator 
\ref{limi} or the $<\bar qq>$ condensate, it is useful to consider 
the theory in an Euclidean space-time box $L\times L\times L\times L$
with periodic boundary conditions and to integrate over quark fields 
first. This leads to a quantum mechanical problem of a single quark in 
a random gluonic background $G^{a}_{\mu}(x)$, which is defined by the 
hermitean Hamiltonian
\begin{equation}
	H=\gamma_{\mu}(\partial_{\mu}+iG^{a}_{\mu}t^{a})\ .
	\label{H1}
\end{equation}
The result of this integration over quarks may be formally expressed
in terms of eigenvalues $\lambda_{n}$ and of (orthonormal) eigenvectors
$\phi_{n}(x)$ of the Dirac Hamiltonian $H$. The spectrum is symmetric
around 0:
$\gamma_{5}\phi_{n}=\phi_{-n}$,\linebreak $ \lambda_{-n}=-\lambda_{n}$. 
Subsequently, the resulting expression has to be averaged over all 
gluon configurations:
\begin{equation}
	<<X[G]>>=\int 
	d[G]\exp(-S_{YM}[G])\prod_{\lambda_{n}>0}(m^{2}+\lambda^{2}_{n})^{N_{f}}X[G]\ .
	\label{XG}
\end{equation}
The fact that the integral \ref{XG} involves a positive probability 
measure suggests a possible analogy with disordered systems.

For the chiral condensate one gets
\begin{equation}
	<\bar qq>=-\lim_{m\to 0}
	\lim_{L\to\infty}\frac{1}{L^{4}}<<\sum_{n}\frac{m}{m^{2}+\lambda^{2}_{n}}>>
	\label{qqbar}
\end{equation}
where $m$ is the quark mass. Similarly, the correlator \ref{limi} is 
given by the formula
\begin{equation}
	F^{2}_{0}=\lim_{m\to 0}
	\lim_{L\to\infty}\frac{1}{L^{4}}<<\sum_{kn}\frac{m}{m^{2}+\lambda^{2}_{k}}
	\frac{m}{m^{2}+\lambda^{2}_{n}}J_{kn}>>\ ,
	\label{F02}
\end{equation}
where
\begin{equation}
	J_{kn}=\frac{1}{4}\sum_{\mu}\vert\int dx 
	\phi^{+}_{k}(x)\gamma_{\mu}\phi_{n}(x)\vert^{2}\ .
	\label{Jkn}
\end{equation}
It is seen that both order parameters \ref{qqbar} and \ref{F02} are 
merely sensitive to the infrared  end of the Dirac spectrum, 
$\vert\lambda_{n}\vert<\epsilon$.

Consider now the Dirac Hamiltonian \ref{H1} as a generator of 
evolution in a fictitious time t added to the 4 Euclidean 
space-coordinates $x_{\mu}$. In this 4+1 dimensional space-time one 
may switch on a homogenous color singlet electric field, adding to $H$ 
a time dependent perturbation
\begin{equation}
	\delta H=i\gamma_{\mu}\xi_{\mu}\frac{\sin\omega t}{\omega}
	\label{deH}
\end{equation}
with $\xi_{\mu}$ constant. $J_{kn}$ is then proportional to the 
$\vert k>\to\vert n>$ transition probability triggered by the 
perturbation \ref{deH} with $\lambda_{k}-\lambda_{n}=\pm\omega$. This 
suggests that from the point of view of the fictitious 4+1 dimensional 
space-time, there is an analogy between $SB\chi S$ and the 
electrically induced transport properties of massless quarks in a 
random medium which is characterized by a coloured magnetic type 
(static) disorder $G^{a}_{\mu}(x)$ with a probability distribution 
given by eq. \ref{XG}.

Indeed, the formula \ref{F02} can be rewritten as
\begin{equation}
	F^2_{0}=\pi^{2}\lim_{\epsilon\to 0}\lim_{L\to\infty}L^{4}\bar 
	J(\epsilon,L)[\rho(\epsilon,L)]^{2}\ ,
	\label{pi2}
\end{equation}
where $\bar J(\epsilon,L)$ is the transition probability \ref{Jkn}
averaged over all initial and final states with "energy"
$\vert\lambda\vert<\epsilon$ and, of course, over the disorder, whereas
$\rho(\epsilon,L)$ stands for the density of states, i.e. the number
of states per unite energy $\epsilon$ and volume $L^{4}$. The latter 
quantity defines the chiral condensate \ref{qqbar} \cite{Banks}:
\begin{equation}
	<\bar qq>=-\pi\lim_{\epsilon\to 0}\lim_{L\to\infty}\rho(\epsilon,L)\ .
	\label{-pi}
\end{equation}
Eq. \ref{pi2} can be viewed as an ultrarelativistic $(m=0)$ version of 
the Kubo-Greenwood formula for electric conductivity (see e.g. 
\cite{Mott}, \cite{Thouless}). It shows that $SB\chi S$ results from 
a conjonction of an appropriate "quark mobility" $\bar J(\epsilon,L)$
and a density of states $\rho(\epsilon,L)$. On the other hand, quark 
condensation is an exclusive affair of the density of states.

The pattern of $SB\chi S$ depends on the degree of accumulation of 
eigenvalues $\lambda_{n}$ near zero, as $L\to \infty$. Suppose that 
the lowest eigenvalues averaged over the disorder behave as 
$<<\lambda_{n}>>\sim L^{-\kappa}$, where $\kappa\geq 1$
as shown by \cite{Vafa84a}. Then for $\epsilon\to 0,\ L\to\infty$
\begin{equation}
	\rho(\epsilon,L)=\mu^{3}\left(\frac{2\epsilon}{\mu}\right)^{\frac{4}
	{\kappa}-1}\ ,
	\label{rho}
\end{equation}
where $\mu$ is a mass scale. Consequently, $\bar qq$ pairs condense 
if and only if $\kappa=4$ \cite{Leutwyler92}. The case 
$\kappa>4$ represents a too strong infrared singularity, which 
can be excluded: $<\bar qq>$ cannot explode \cite{Gasser97a}. On the 
other hand, $SB\chi S$, i.e. a non-zero value of $F^{2}_{0}$ given 
by Eq. \ref{pi2} can be shown to require $\kappa\geq 2$. Hence, 
one is faced to two extreme alternatives of $SB\chi S$:

{\bf i)}  \boldmath$\kappa=2$\unboldmath: The density of states and $<\bar qq>$ 
vanish as $\epsilon$, but still $F^{2}_{0}\neq 0$, i.e. $SB\chi S$ 
takes place and Goldstone bosons are formed, due to a large mobility 
of "low-energy" quarks, $\bar J\sim\epsilon^{-2}L^{-4}$. This behavior 
occurs naturally provided quark states are delocalized.

{\bf ii)}  \boldmath$\kappa=4$\unboldmath: The density of states remains non-zero 
as $\epsilon\to 0$, i.e.\linebreak $<\bar qq>\neq 0$, whereas the mobility 
$\bar J$ must be suppressed by a factor $L^{-4}$. Such a suppression 
of mobility could be naturally understood if the Euclidean quark 
states were in a sense localized.

The intermediate case $2<\kappa<4$ cannot be excluded on general 
grounds. One can show, however, that an effective low energy theory 
characterized by an effective Lagrangian analytic in quark masses can 
only exist, provided $4/\kappa=$ integer. This, together with 
the condition $\kappa\geq 2$, selects the cases
$\kappa=2$ and $\kappa=4$ as two distinct possibilities of 
realizing $SB\chi S$ in QCD.

In Nature, both types of states belonging to the $\kappa=2$ and
$\kappa=4$ bands can coexist and contribute to the $SB\chi S$. 
Since only the $\kappa=4$ band contributes to the chiral 
condensate, the actual value of $<\bar qq>$ can hardly be guessed in 
advance. What matters in practice, is the size of the parameter
\begin{equation}
	B_{0}=-\frac{1}{F^{2}_{0}}<\bar qq>\ .
	\label{bo}
\end{equation}
One will have to distinguish on phenomenological grounds between a 
{\bf large condensate}, typically $B_{0}$ (1 GeV) $\sim$ 2 GeV which 
seems to be suggested by lattice simulations and a {\bf small 
condensate}, say $B_{0}\sim$ 100 MeV resulting from an attempt to 
variationally extend the QCD perturbation theory \cite{Neveu} and to 
calculate non-perturbative quantities such as $F_{0}$ and $<\bar qq>$
\cite{Kneur}.
\section{Quark mass ratios}
We consider the 3 light quark masses $m_{u}(\mu),\ m_{d}(\mu)$ and 
$m_{s}(\mu)$ renormalized in the $\overline{MS}$ scheme at the running 
scale $\mu$. In this section we will be mostly concerned with the 
relation between the two quark-mass ratios
\begin{equation}
	r=\frac{m_{s}}{\hat m},\ R=\frac{m_{s}-\hat m}{m_{d}-m_{u}},
	\ \hat m=\frac{1}{2}(m_{u}+m_{d})
	\label{rms}
\end{equation}
and the masses of unmixed Goldstone bosons $\pi^{+},\ K^{+}$ and $K^{0}$.
\subsection{The standard picture}
The generally accepted picture of the ratios \ref{rms} goes back to 
\cite{Weinberg77} and has been further elaborated by Gasser and 
Leutwyler \cite{Gasser82}, \cite{Gasser85}, 
\cite{Leutwyler90},\cite{Leutwyler96}. One writes
\begin{equation}
	\begin{array}{rcl}
		M^{2}_{\pi^{+}} & = & (m_{u}+m_{d})B_{0}+\Delta_{\pi^{+}}  \\
		M^{2}_{K^{+}} & = & (m_{u}+m_{s})B_{0}+\Delta_{K^{+}}    \\
		M^{2}_{K^{0}} & = & (m_{d}+m_{s})B_{0}+\Delta_{K^{0}}
	\end{array}
	\label{3m}
\end{equation}
and one assumes that
\begin{equation}
	\epsilon_{P}=\frac{\Delta_{P}}{M^{2}_{P}}<<1\ .
	\label{ep}
\end{equation}
It follows \cite{Weinberg77}
\begin{equation}
	r=r_{2}\{1+O(\epsilon_{P})\},\ 
	r_{2}=2\frac{M^{2}_{K}}{M^{2}_{\pi}}-1\simeq 25.9
	\label{rr2}
\end{equation}
and
\begin{equation}
	R\frac{\Delta M^{2}_{K}}{M^{2}_{K}-M^{2}_{\pi}}=1+O(\epsilon_{P}),\ 
	\Delta M^{2}_{K}=(M^{2}_{K^{0}}-M^{2}_{K^{+}})_{QCD}\ .
	\label{RD}
\end{equation}
Gasser and Leutwyler have shown \cite{Gasser85} that the 
$O(\epsilon_{P})$ corrections in Eqs. \ref{rr2} and \ref{RD} are 
related. Eliminating them one gets a hyperbolic relation
\begin{equation}
	R\frac{\Delta M^{2}_{K}}{M^{2}_{K}-M^{2}_{\pi}}=\frac{r_{2}+1}{r+1}
	\{1+O(\epsilon^{2}_{P})\}\ ,
	\label{Rr1}
\end{equation}
which is (almost exactly) equivalent to the Leutwyler's ellipse 
\cite{Leutwyler90}
\begin{equation}
	\left(
	\frac{m_{u}}{m_{d}}\right)^{2}+\frac{1}{Q^{2}}
	\left(
	\frac{m_{s}}{m_{d}}\right)^{2}=1+O(\epsilon^{2}_{P}),\ 
	Q^{2}=\frac{M^{2}_{K}-M^{2}_{\pi}}{\Delta 
	M^{2}_{K}}\frac{M^{2}_{K}}{M^{2}_{\pi}}\ .
	\label{mumd}
\end{equation}
This constitutes what is usually called "a current-algebra picture 
of quark mass ratios" despite the fact that the assumption \ref{ep} 
follows neither from current algebra nor from QCD. None of the 
conclusions \ref{rr2}, \ref{RD}, \ref{Rr1} or \ref{mumd} holds if 
$\epsilon_{P}\sim 1$, in particular, if $B_{0}\sim 0$.
\subsection{Expansion of Goldstone boson masses}
$\epsilon_{P}\sim 1$ does not mean that the expansion of $M^{2}_{P}$ 
in powers of $m_{q}$ breaks down. One can actually argue that 
expansion of $\Delta_{P}$ proceeds by powers of $m_{q}/\Lambda_{H}$, 
where $\Lambda_{H}\sim$ 1 GeV is the characteristic mass-scale of 
first massive bound states. (The same parameter $m_{q}/4\pi F_{0}$ 
characterizes the coefficients of chiral logs, $4\pi 
F_{0}\sim\Lambda_{H}$).

Writing the expansion of $\Delta_{P}$ as
\begin{equation}
	\Delta_{P}=\frac{1}{F^{2}_{0}}\sum_{n=2}^{\infty}a_{n}m^n_{q}+\mbox{(Chiral 
	log's)},
	\label{DP}
\end{equation}
the coefficients $a_{n}$ represent connected n-point functions of scalar 
and pseudoscalar quark bilinears with all Goldstone boson poles 
subtracted at vanishing external momenta. They are dominated by 
exchanges of bound states of mass $\sim\Lambda_{H}\sim$ 1 GeV and 
this fact allows one to roughly estimate their order of magnitude 
\cite{Georgi}:
\begin{equation}
	a_{n}=c_{n}F^{2}_{0}\Lambda^{2-n}_{H},\ c_{n}\sim 1\ .
	\label{an}
\end{equation}
The coefficients $a_{n}$ are genuine QCD quantities independent of the 
Goldstone boson sector of the theory and the estimate \ref{an} should 
hold independently of the size of the chiral condensate. On the other 
hand, the parameters $\epsilon_{P}$ measure the relative importance 
of the condensate and $\Delta_{P}$ contributions to $M^{2}_{P}$. 
Since $a_{2}$ is of the order $F^{2}_{0}$, one can expect 
$\epsilon_{P}\sim m_{q}/(m_{0}+m_{q})$, where $m_{0}\sim B_{0}$. If 
$B_{0}$ happened to be as small as, say, $m_{s}$, the assumption 
$\epsilon_{K}\ll 1$ would break down. Hence, according to the size 
of $B_{0}$ as compared to $\Lambda_{H}$, one should distinguish two 
expansion schemes of the effective Lagrangian and of $M^{2}_{P}$:

{\bf i) Standard Chiral Perturbation Theory} - $S\chi PT$ 
\cite{Gasser84, Gasser85, Gasser97} is a simultaneous expansion in powers of 
$m_{q}/\Lambda_{H}$ and $m_{q}/m_{0}$. It is defined by the chiral 
counting $m_{q}=O(p^{2}),\ B_{0}\sim m_{0}\sim\Lambda_{H}=O(1)$.

{\bf ii) Generalized Chiral Perturbation Theory} - $G\chi PT$ 
\cite{Fuchs91},\linebreak \cite{Stern93}, \cite{Knecht95a} is an expansion in 
$m_{q}/\Lambda_{H}$ alone. ($m_{q}\ll m_{0}$ is {\bf not} assumed). 
This can be realized systematically, through a modified chiral 
counting: $m_{q}=O(p),\ m_{0}\sim B_{0}=O(p)$. At any given order 
$O(p^n)$, $S\chi PT$ appears as a special case of $G\chi PT$.
\subsection{The ratio $r=m_{s}/\widehat m$ in $G\chi PT$}
Let us put, for a moment, $m_{u}=m_{d}=\widehat m$ and write the 
expansion of $F^{2}_{\pi}M^{2}_{\pi}$ and $F^{2}_{K}M^{2}_{K}$ in the 
form
\begin{equation}
	\begin{array}{lcl}
		\frac{1}{F^{2}_{0}}F^{2}_{\pi}M^{2}_{\pi} & = & 2\widehat m 
		B(\mu)+4\widehat m^{2}A_{0}(\mu)+
		\frac{1}{F^{2}_{0}}F^{2}_{\pi}\delta M^{2}_{\pi}  \\
		\frac{1}{F^{2}_{0}}F^{2}_{K}M^{2}_{K} & = & (m_{s}+\widehat m)
		B(\mu)+(m_{s}+\widehat m)^{2}A_{0}(\mu)+
		\frac{1}{F^{2}_{0}}F^{2}_{K}\delta M^{2}_{K}\ , 
	\end{array}
	\label{ms}
\end{equation}
which, as it stands, holds independently of the size of
\begin{equation}
	B(\mu)=B_{0}+2(m_{s}+2\widehat m)Z_{0s}(\mu)\ .
	\label{Bmu}
\end{equation}
$A_{0}(\mu)$ and $Z_{0s}(\mu)$ are related to two-point functions of 
scalar and pseudoscalar currents. $Z_{0s}(\mu)$ violates the Zweig 
rule (it is suppressed as\linebreak $N_{c}\to\infty$) and one may expect 
$B(\mu)\simeq B_{0}$. Both $A_{0}(\mu)$ and $Z_{0s}(\mu)$ are 
constants of the effective Lagrangian renormalized at the scale 
$\mu$. $\delta M^{2}_{P}$ involves chiral logs and higher order terms. 
They are relatively small: $\delta M^{2}_{P}<0.1 M^{2}_{P}$ 
independently of the magnitude of $B$.

$S\chi PT$ considers the first term in Eq. \ref{ms} as the leading 
$O(p^{2})$ contribution and the remaining two terms as a small 
$O(p^{4}),\ O(p^{6})\ldots$ perturbation. $G\chi PT$ admits that the 
first two terms could equally contribute to the leading $O(p^{2})$ 
order, (because $B$ is small) and it treats $\delta M^{2}_{P}$ as a 
small $O(p^{3}),\ O(p^{4})\ldots$ perturbation. It is clear that 
$S\chi PT$ can be applied only provided
\begin{equation}
	\widehat m\ll m_{0}=\frac{B_{0}}{2A_{0}},\ m_{s}\ll 2m_{0}\ .
	\label{mm0}
\end{equation}
If the condition \ref{mm0} is not satisfied, Eq. \ref{ms} does not 
allow to determine the ratio $r=m_{s}/\widehat m$ as in Eq. \ref{rr2} . At 
most can one relate the magnitude of $r$ to the amount of violation of 
the GOR relation \cite{Gellmann}, $M^{2}_{\pi}=2\widehat m B_{0}$.
Eqs. \ref{ms} yield the exact formula
\begin{equation}
	X_{GOR}\equiv\frac{2\widehat mF^{2}_{0}B}{F^{2}_{\pi}M^{2}_{\pi}}=
	\frac{1}{r^{2}-1}
	\{r-r^{*}_{1}(r)\}
	\{r+r^{*}_{1}(r)+2\}
	\left(\frac{M^{*}_{\pi}}{M_{\pi}}\right)^{2}
	\label{GOR}
\end{equation}
where
\begin{equation}
	r^{*}_{1}(r)=2\frac{F_{K}}{F_{\pi}}\ 
	\frac{M^{*}_{K}(r)}{M^{*}_{\pi}(r)}-1\ ,\quad
	[M^{*}_{P}(r)]^{2}=M^{2}_{P}-\delta M^{2}_{P}(r)\ .
	\label{ret}
\end{equation}
The dependence of the GOR-ratio $X_{GOR}$ on the quark mass ratio $r$ 
is displayed in Fig. 1, together with the uncertainties attached with 
our estimates of higher order corrections. Eq. \ref{GOR} meets a zero 
at $r=r_{crit}$, defined by
\begin{equation}
	r_{crit}=r^{*}_{1}(r_{crit})=2\frac{M_{K}}{M_{\pi}}-1+O(m)=6.3+O(m)\ .
	\label{crit}
\end{equation}
Including higher order effects, one gets
\begin{equation}
	r_{crit}=8.22\pm 0.64\ .
	\label{crit2}
\end{equation}
Since vacuum stability requires $B\geq 0$, one must have $r\geq r_{crit}$. 
Smaller $B$, closer the ratio $r$ to its critical value \ref{crit2}.
\begin{center}
\includegraphics*[width=10cm]{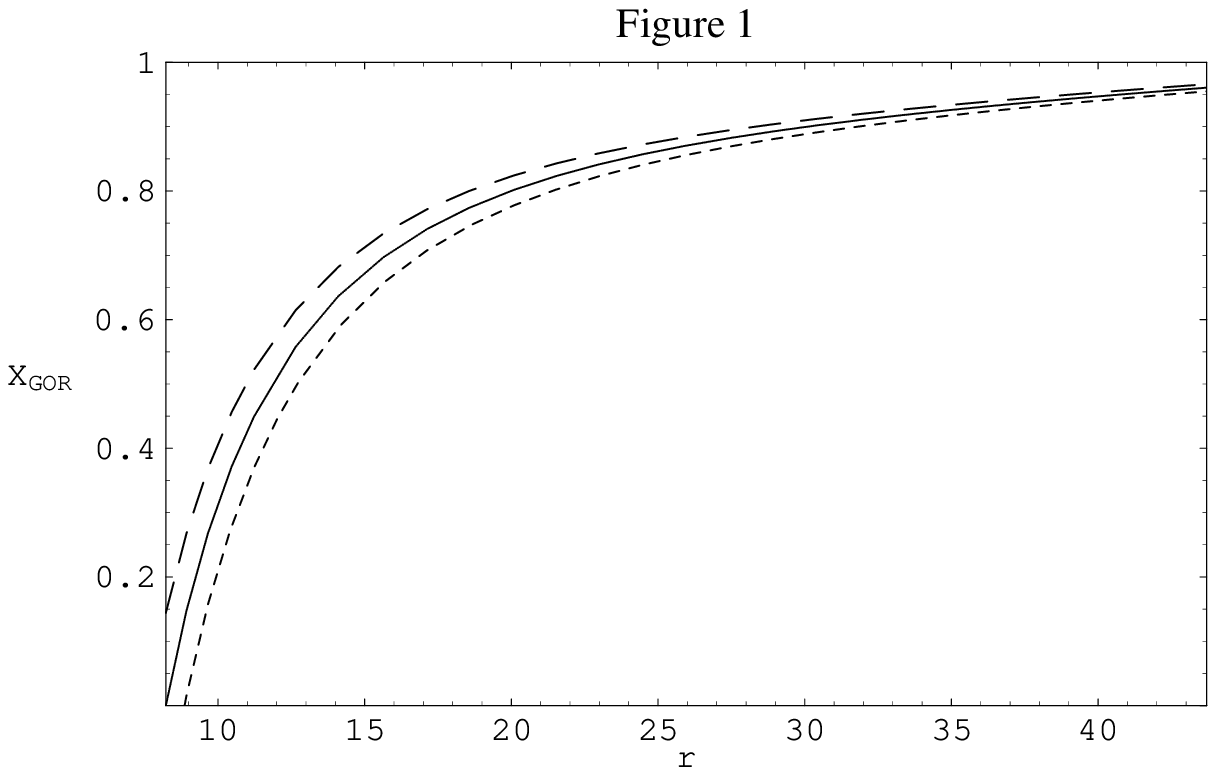}
\end{center}
\begin{center}
{\baselineskip 12 pt Fig. 1:  The GOR ratio \ref{GOR} as a function 
of $r=m_{s}/\widehat m$}.
\end{center}
\subsection{The ratio $R=\frac{\textstyle m_{s}-\widehat m}
{\textstyle m_{d}-m_{u}}$ 
in $G\chi PT$: dismiss of the ellipse}
We now turn to the isospin breaking effects due to $m_{d}\neq 
m_{u}$. There is a "magic" {\bf linear combination} of the three 
unmixed Goldstone boson masses \cite{Fuchs91}, \cite{Knecht95b}
\begin{equation}
	R\Delta M^{2}_{K}-(M^{2}_{K}-M^{2}_{\pi})-
	(M^{2}_{K}-\frac{r+1}{2}M^{2}_{\pi})\frac{r-1}{r+1}=O
	\left(\frac{m^{3}_{q}}{\Lambda_{H}}\right)
	\label{O1}
\end{equation}
in which all $O(m),\ O(m^{2})$ and even $O(m^{2}\ln m)$ terms cancel, 
independently of the size of $<\bar qq>$. In order to see the 
connection with Gasser-Leutwylers hyperbola \ref{Rr1}-alias Leutwyler's 
ellipse \ref{mumd}, let us rewrite Eq. \ref{O1}, including the $O(m^{3}_{q})$ 
corrections, as
\begin{equation}
\begin{array}{rcl}
	R\frac{\Delta M^{2}_{K}}{M^{2}_{K}-M^{2}_{\pi}}&=&
	{\textstyle\frac{r_{2}+1}{r+1}-
	\frac{(r_{2}-r)^{2}}{(r+1)(r_{2}-1)}}-\\
	&-&\left\{
	\left(\frac{F^{2}_{K}}{F^{2}_{\pi}}-1\right)\frac{r_{2}-r}{r_{2}-1}+
	\frac{m^{3}_{s}\rho_{2}}{M^{2}_{K}-M^{2}_{\pi}}\frac{(r-1)^{2}(r+1)}{r^{3}}
	+\cdots\right\}
	\end{array}
	\label{r+1}
\end{equation}
where $r_{2}=2M^{2}_{K}/M^{2}_{\pi}-1\simeq 25.9$ and 
$\vert\rho_{2}\vert\lesssim 1/\Lambda_{H}$ is a constant 
contained in the $\mathfrak{L}_{03}$ component of $\mathfrak{L}_{eff}$ 
\cite{Knecht95a, Knecht95b}. Keeping only the first term on the right hand side 
of Eq. \ref{r+1} corresponds to the Gasser-Leutwyler hyperbola (shown 
as the dashed curve on Fig. 2). Including also the second term (which in 
$S\chi PT$ would be $O(m^{2})$), one obtains the leading order $G\chi PT$ 
expression, which differs from the standard case to the extent $r$ 
differs from $r_{2}$, (see the solid line on Fig. 2). Finally, the curly 
bracket represents the NLO $G\chi PT$ correction, which is included 
in the long dashed curve on Fig. 2 for the case $\rho_{2}=0$. Fig. 3 shows the 
details of the whole expression \ref{r+1}, including the uncertainty 
in $\rho_{2}$.

The leading order $S\chi PT$ formula \ref{RD} implies $R\simeq 43$, 
assuming the validity of the Dashen theorem in evaluating $\Delta 
M^{2}_{K}$, \cite{Gasser82}. For lower $r$, say $r\sim 10$, $G\chi PT$  
suggests a similar value, typically $R\simeq 40 \div 45$, since the 
10-20\% increase of the ratio \ref{r+1} is compensated by a similar 
increase of $\Delta M^{2}_{K}$, due to the violation of the Dashen 
theorem, \cite{Bijnens93}, \cite{Moussallam97}. 

\vspace{1cm}
\begin{minipage}[c]{.70\linewidth}
\includegraphics*[width=8cm]{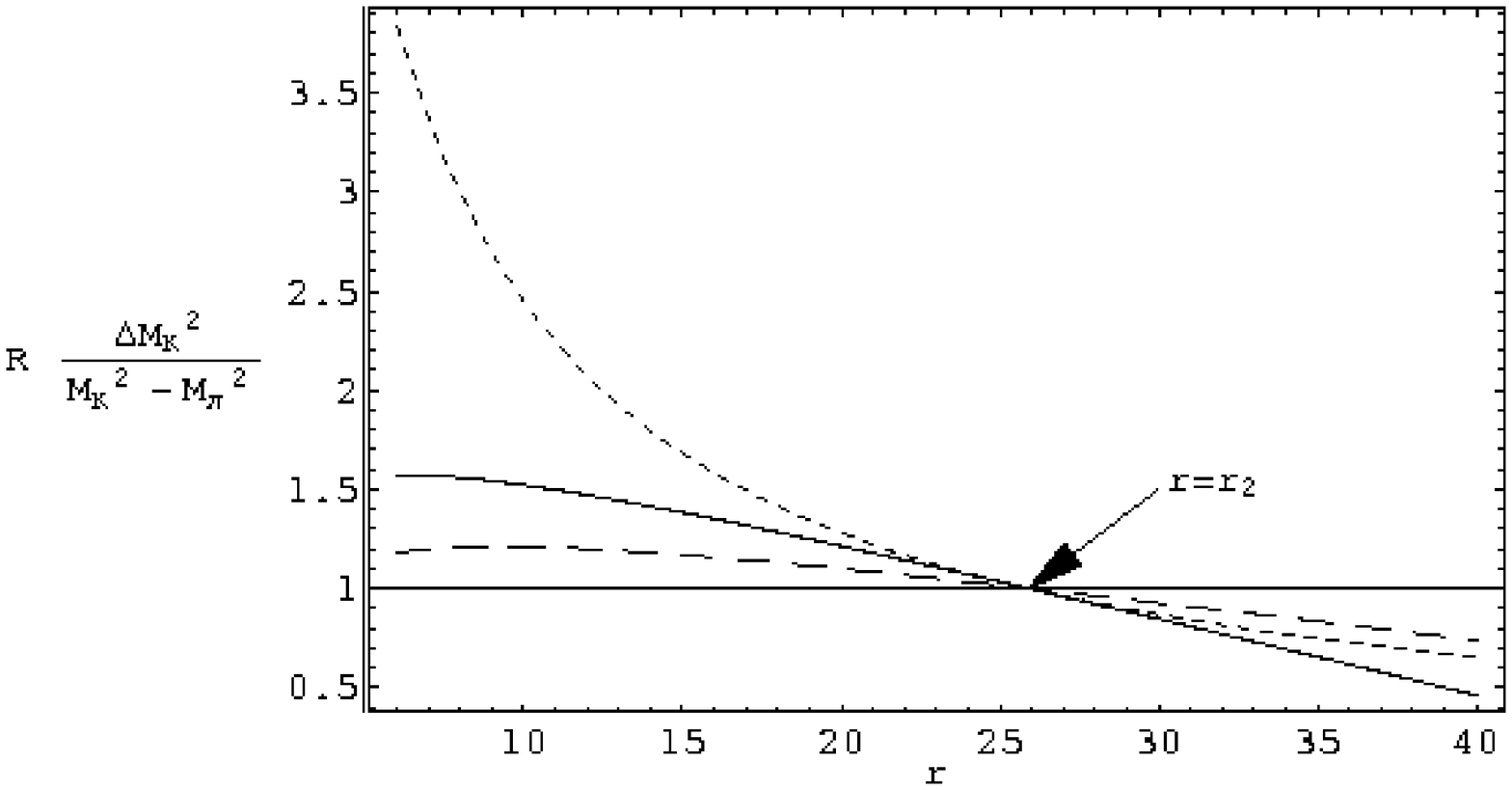}
\end{minipage}\hfill
%\end{figure}
\begin{minipage}[c]{.27\linewidth}
Fig. 2: Relation between the two quark mass ratio $R$ and $r$,
(see the text after Eq. \ref{r+1}
\end{minipage}

\vspace {1cm}
\begin{minipage}[c]{.27\linewidth}
Fig. 3: Sensitivity of the relation between $R$ and $r$ to the NLO parameter
$\rho_2$. $\rho_{2}=0$ solid curve, $\rho_{2}=\pm\frac{1}{M_\rho}$ dashed
and long-dashed curves.
\end{minipage}\hfill
%\begin{figure}
\begin{minipage}[c]{.70\linewidth}
\includegraphics*[scale=0.33]{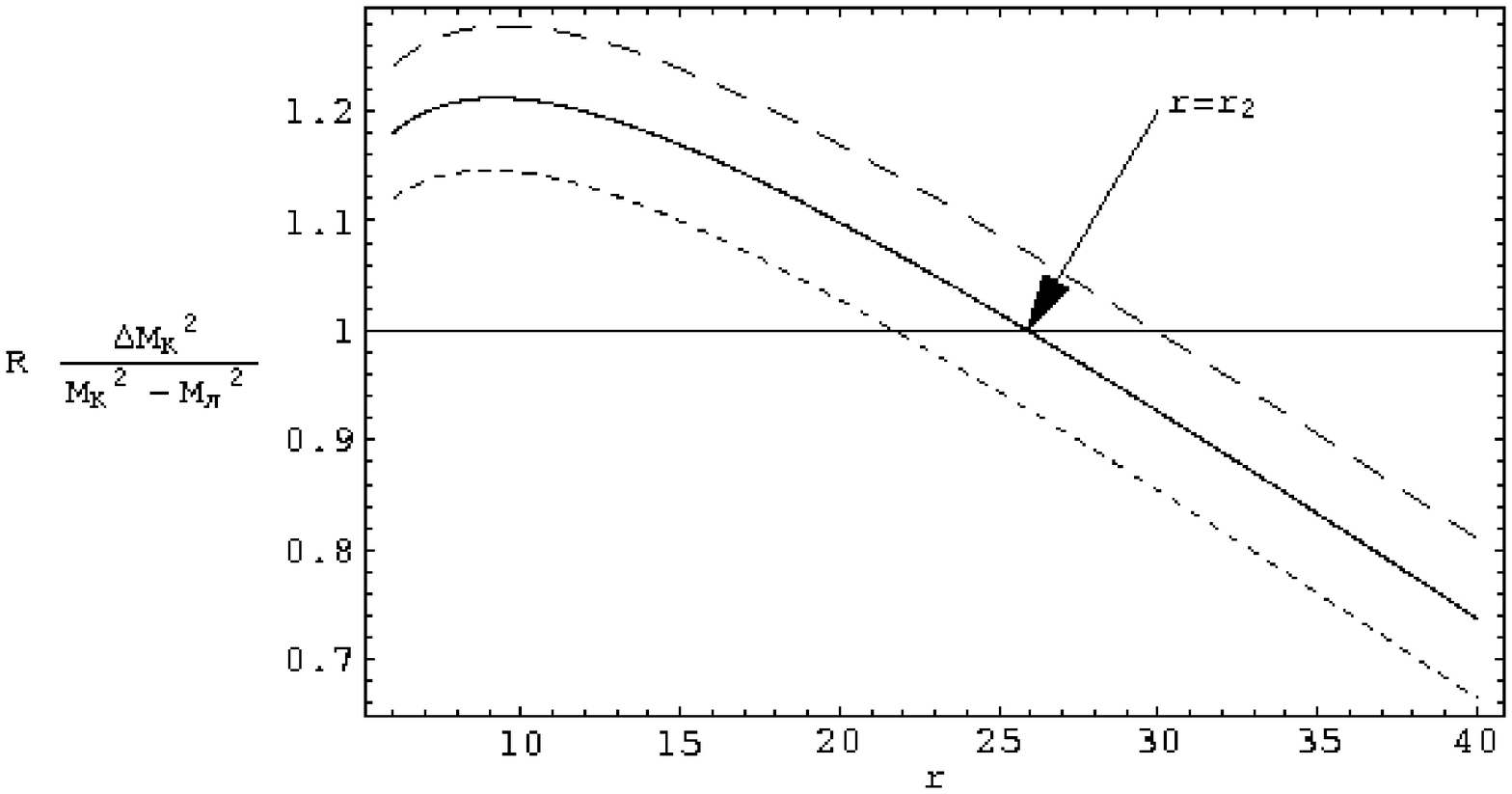}
\end{minipage}
%\end{figure}
\section{The value of the strange quark mass}
Recently, the first experimental determination of $m_{s}(\mu)$ has
been reported \cite{Chen97}, based on a precise measurement by ALEPH 
collaboration of inclusive branching ratio's $R^{S=0}_{\tau}$ and 
$R^{S=1}_{\tau}$ of $\tau$-lepton into $\nu_{\tau}$ and hadronic final states 
with total strangeness 0 and 1 respectively \cite{Davier}. The 
determination follows the method of Braaten, Narison and Pich 
\cite{Braaten} which, using the QCD OPE at the scale $M^{2}_{\tau}$, 
gives an expression of $R^{S}_{\tau}$ in terms of $\alpha_{s}(M_{\tau}),\ 
m_{s}(M_{\tau})$ and various (less important) non-perturbative parameters. 
The preliminary result reads \cite{Chen97}
\begin{equation}
	m_{s}(M_{\tau})=
	\left(172^{+26}_{-31}\right)\ \mbox{MeV},\ 
	m_{s}(\mbox{1 GeV})=
	\left(235^{+35}_{-42}\right)\ \mbox{MeV}\ .
	\label{mtau}
\end{equation}
This value is somewhat higher than expected from recent lattice 
\cite{LATT} and some sum rule \cite{SR} estimates. It is, however, compatible 
with a former determination by \cite{Narison}, which is based on a 
similar method, but uses less accurate $e^{+}e^-$ data as input.
\subsection{Consequences for parameters $B$ and $A_{0}$}
It has been already shown that Eqs. \ref{ms} lead to the expression 
of $\widehat mB(\mu)$, Eq. \ref{GOR}, as a function of 
$r=m_{s}/\widehat m$. A similar expression can be obtained for 
$\widehat m^{2}A_{0}$:
\begin{equation}
	\frac{4\widehat m^{2}F^{2}_{0}A_{0}}{(F_{\pi}M^{*}_{\pi})^{2}}=
	2\frac{r^{*}_{2}(r)-r}{r^{2}-1}\ ,
	\label{4m}
\end{equation}
where
\begin{equation}
	r^{*}_{2}(r)=2\left[
	\frac{F_{K}M^{*}_{K}(r)}{F_{\pi}M^{*}_{\pi}(r)}\right]^{2}-1\ .
	\label{2F}
\end{equation}

\begin{minipage}[c]{.70\linewidth}
\includegraphics*[width=8cm]{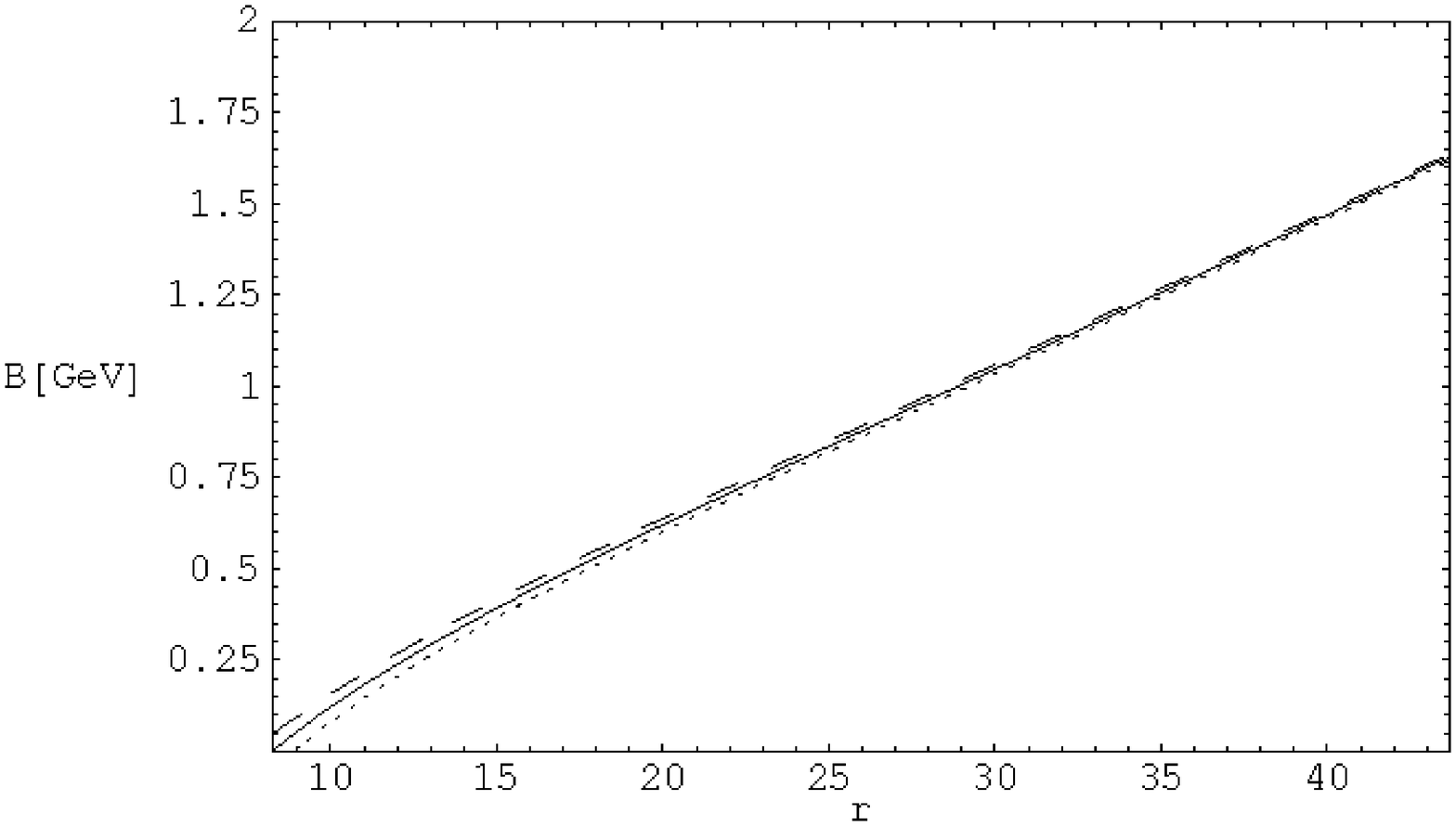}
\end{minipage}\hfill
\begin{minipage}[c]{.27\linewidth}
Fig. 4: The condensate parameter B \ref{Bmu}
as a function of $r=m_{s}/\widehat m$.
\end{minipage}
%\vspace{0.5cm}
Using the central value $m_{s}$ (1 GeV) = 235 MeV, one can obtain the
condensate parameter $B(\mu)$ and the quadratic slope parameter
$A_{0}(\mu)$ as functions of $r$. The results are displayed on Figs. 4
and 5. They correspond to the QCD scale $\nu=1$ GeV and to the
$\chi PT$ scale $\mu= M_{\eta}$. One observes that for lower values
of $r$, the critical mass scale $m_{0}=B_{0}/2A_{0}$ can indeed be
rather small: $m_{0}\sim$ (20-25) MeV for $r\sim 10$. Under these
circumstances, even the non-strange quark mass $\widehat m\sim$ 235
MeV/r would be comparable to $m_{0}$, invalidating the use of the
$S\chi PT$ even in the non-strange sector.
%\begin{figure}
%\vspace{0.3cm}
\begin{minipage}[c]{.27\linewidth}
Fig. 5: The parameter $A_{0}$ (at $\mu=M_\eta$) as a function of $r$. The NLO
uncertainty is shown as in Figure 3.
\end{minipage}\hfill
\begin{minipage}[c]{.70\linewidth}
\includegraphics*[scale=0.35]{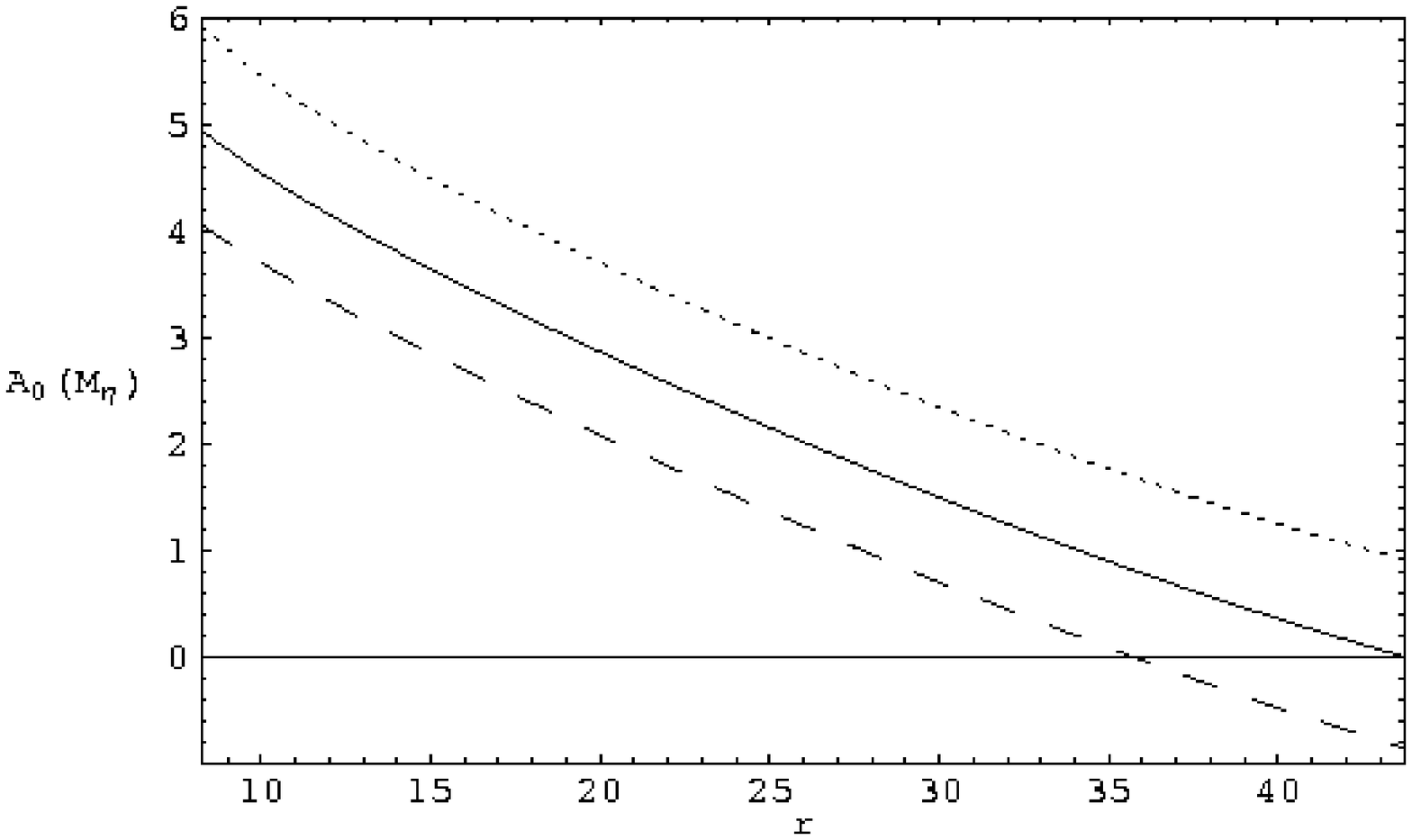}
\end{minipage}
%\end{figure}
\subsection{Natural size of $A_{0}$}
Let us consider the two-point function of (octet) scalar and 
pseudoscalar quark currents $S^{a}(x)$ and $P^{a}(x)$ respectively. In 
the chiral limit $m_{s}=\widehat m=0$, one has for small $q^{2}$
\begin{equation}
\begin{array}{l}
	\prod_{SP}(q^{2})\delta^{ab}  =  \frac{i}{F^{2}_{0}}
	\int dx \ e^{iqx}<0\vert
	\{S^{a}(x)S^b(0)-P^{a}(x)P^b(0)\}\vert 0>= \\
	  =  \delta^{ab}
	 \left\{
	 \frac{B^{2}_{0}}{q^{2}}+\frac{5}{96\pi^{2}}\frac{B^{2}_{0}}{F^{2}_{0}}
	 [\ln\frac{\mu^{2}}{-q^{2}}+1]+A_{0}(\mu)+O(q^{2})
	 \right\}\ .
	\label{SP}
\end{array}
\end{equation}
The first and second terms on the right-hand side represent the GB 
pole and loop respectively, whereas the constant $A_{0}(\mu)$ receives 
contributions from exchanges of {\bf massive} $O^{++}$ and $O^{-+}$ 
resonances: $f_{0},\ \pi\prime\ldots$. This fact suggests that the 
constant $A_{0}(\mu)$ should be rather insensitive to the GB sector of 
the theory , in particular, to the size of $B_{0}$. $A_{0}(\mu)$ is 
related to the renormalized $S\chi PT$, $O(p^{4})$ low-energy 
constants $L^r_{8}(\mu)$,
\begin{equation}
	L^r_{8}(\mu)=\frac{F^{2}_{0}}{16B^{2}_{0}}A_{0}(\mu)\ ,
	\label{LR}
\end{equation}
which has been estimated \cite{Gasser85} to be 
$L^r_{8}(M_{\eta})=(1.1\pm 0.3)\times 10^{-3}$. Taking the standard 
value of the condensate, $B_{0}\simeq 1.5$ GeV (at QCD scale 
$\nu=1$ GeV), this corresponds to
\begin{equation}
	A_{0}(M_{\eta})_{S\chi PT}=4.5\pm 1.2\ .
	\label{Meta}
\end{equation}
Notice that this estimate, which is obtained assuming a large value 
of the condensate, is comparable to the values displayed on Fig. 5 for 
the case of lower $r$ $(r\sim 10)$, for which the condensate 
$B_{0}\sim B$ is about ten times smaller, c.f. Fig. 4. We believe that 
this is not accidental: on the one hand, there are quantities like 
$L_{8}$, or the matrix element $<0\vert P^{a}\vert\pi^b>$, which 
are rather sensitive to the size of $B_{0}$, because they either 
diverge or vanish as $B_{0}\to 0$. On the other hand, couplings and 
masses of non-Goldstone particles $(f_{0},\pi\prime\ldots)$ and, 
consequently, the constant $A_{0}$ show a moderate dependence on the 
chiral condensate $<\bar qq>$.
\section{QCD sum rules}
If $m_{s}\simeq 235$ MeV and $r\simeq 10$, the non-strange mass 
$\widehat m$ should be $3\div 4$ times larger than the typical 
output of existing standard QCD sum rule analysis 
\cite{Dominguez}, \cite{Bijnens95}. In this section it will be 
argued that whereas the method of QCD sum rules is by itself 
perfectly adapted for a determination of quark masses and 
condensates, the crucial experimental data needed as input are still 
missing. Instead, in existing determinations of $\widehat m$, data are 
replaced by models which implicitly assume a large value of $<\bar qq>$. 
Consequently, the resulting analysis represents a consistency check of 
the large condensate scenario, rather than an independent 
determination of $\widehat m$.
\subsection{Two-point function of $\partial^\mu A_{\mu}$}
I will mainly concentrate on the most elaborated and relevant example of 
the two point function of $D^{\bar ud}_{5}\equiv \partial^\mu[\bar 
u\gamma_{\mu}\gamma_{5}d]=2\widehat m\bar ui\gamma_{5}d$:
\begin{equation}
	\psi^{\bar ud}_{5}(q^{2})=i
	\int dx e^{iqx}<\Omega\vert TD^{\bar ud}_{5}(x)D^{\bar 
	du}_{5}(0)\vert\Omega>\ .
	\label{psi}
\end{equation}
The corresponding imaginary part can be written as
\begin{equation}
	\begin{array}{rcl}
		\frac{1}{\pi}Im\psi^{\bar ud}_{5}(t) & = & 
		2F^{2}_{\pi}M^{4}_{\pi}\delta(t-M^{2}_{\pi})+\rho(t)  \\
		\rho(t) & = & \rho_{3\pi}(t)+\rho_{K\bar K\pi}(t)+\ldots
	\end{array}
	\label{Im}
\end{equation}
The spectral function $\rho(t)$ is in principle measurable in tau 
decays, but sofar, neither its normalization nor its shape are 
known: $\rho(t)$ is proportional to $\widehat m^{2}$ and it is 
precisely $\widehat m$ we want to determine. Three informations are 
available: {\bf i)} For large t, the QCD asymptotics takes over,
\begin{equation}
	\rho(t)\longrightarrow\frac{3}{2\pi^{2}}[\widehat m(t)]^{2}t
	\left\{1+\frac{17}{3}\frac{\alpha_{s}(t)}{\pi}+\ldots\right\}\ .
	\label{rhot}
\end{equation}
{\bf ii)} In the intermediate energy region, there are two 
resonances $\pi^{\prime}(1300)$ and $\pi^{\prime\prime}(1770)$ with the couplings to 
the axial current $F_{\pi^{\prime}}$ and $F_{\pi^{\prime\prime}}\sim \widehat m$ and 
unknown. {\bf iii)} Finally, for $t\sim 0$, the dominant component 
$\rho_{3\pi}(t)$ is given by $G\chi PT$: to the leading order one 
gets \cite{Stern94}
\begin{equation}
	\rho_{3\pi}(t)=\frac{F^{2}_{\pi}}{768\pi^{4}}
	\left(\frac{M_{\pi}}{F_{\pi}}\right)^{4}t
	\left\{1+10\frac{r_{2}-r}{r^{2}-1}+30
	\left(\frac{r_{2}-r}{r^{2}-1}\right)^{2}\right\}+\ldots
	\label{rho3}
\end{equation}
as a function of $r=m_{s}/\widehat m$. The special case of $S\chi PT$, 
i.e. of the large condensate, corresponds to $r=r_{2}$ and it has been 
suggested to pin down the normalization of $\rho$ using this 
information. This then yields the well known result $\widehat m$ 
(1 GeV) $\sim$ (6-7) MeV, \cite{Dominguez}, \cite{Bijnens95}. The 
problem is that for $B_{0}\sim 0$, one has $r\simeq 
r_{1}=2\frac{M_{K}}{M_{\pi}}-1$ and the curly bracket in Eq. 
\ref{rho3} introduces an enhancement factor 
$1+10\times\frac{1}{2}+30\times\frac{1}{4}=13.5$. This is compatible 
with the expectation that for $B_{0}\sim 0$, the normalization of 
$\rho(t)$ should increase proportionally to the increase of $\widehat 
m^{2}$.

It is worth noting that the positivity of the spectral 
function $\rho(t)$ alone implies interesting {\bf lower bounds for} 
\boldmath$\widehat m\ $\unboldmath  
which are independent of the size of $B_{0}$. 
For a recent discussion of these bounds and for a comparison with 
lattice determinations of $\widehat m$ and $m_{s}$, see \cite{Lellouch}.
\subsection{QCD - Hadron duality}
It has been suggested \cite{Bijnens95}, \cite{Prades} that the 
principle of QCD - hadron duality could help in fixing the 
normalization of $\rho$. One considers the ratio
\begin{equation}
	R_{Had}(s)=\frac{3}{2s}\frac{\int^s_{0}dt\ t\ \mbox{Im}\psi_{5}(t)}
	{\int^s_{0}dt\ \mbox{Im}\psi_{5}(t)}=\frac{3}{2s}
	\frac{\int^s_{0}dt\ t\rho(t)}{2F^{2}_{\pi}M^{4}_{\pi}+
	\int^s_{0}dt\ \rho(t)}
	\label{Rhad}
\end{equation}
and one requires that in a suitable interval of $s$ it coincides with 
a similar ratio $R_{\mathrm{QCD}}(s)=1+$ OPE corrections, calculated in QCD 
perturbation theory. Indeed, within the $S\chi PT$, one expects both 
the pion and the continuum contributions on RHS of Eq. \ref{Rhad} to 
be of a comparable size $O(p^{4})$. Under these circumstances, one 
could expect that $R_{Had}(s)$ will be sensitive to the normalization 
of $\rho$, whereas $R_{\mathrm{QCD}}(s)$ should be almost independent of 
$B_{0}$ and/or $\widehat m$. However, for large enough $\rho$, such 
that in the denominator of \ref{Rhad} the continuum dominates over 
the pion contribution, $R_{{Had}}(s)$ becomes independent of the 
normalization. This is what is {\bf expected} to happen as a 
consequence of the $G\chi PT$ chiral counting: the 
$\rho$-contributions, both in the numerator and denominator of Eq. 
\ref{Rhad} are $O(m^{2})=O(p^{2})$, whereas the pion term is $O(p^{4})$. 
Hence, within the $G\chi PT$ scenario, the duality criterion at most 
constrains the shape of $\rho(t)$ and it should be satisfied {\bf 
without including the pion contribution}. After all, in the scalar 
channel there is no pion pole and $R_{\mathrm{QCD}}(s)$ is about the same 
both in the $O^{++}$ and $O^{-+}$ channels. An explicit counter example to 
the statement that duality constrains the normalization of $\rho$ is 
shown in Figs. 6 and 7. The model of the spectral function in the 
intermediate energy region is displayed in Fig. 6. The corresponding 
ratio $R_{{Had}}(s)$ with no pion contribution included is compared 
in Fig. 7 with $R_{\mathrm{QCD}}(s)$. Notice that due to the (unknown) 
contribution of direct instantons, one should not expect in this 
channel the onset of QCD(OPE)-hadron duality at too low~$s$.
%\begin{figure}
\begin{minipage}[c]{.65\linewidth}
\includegraphics*[width=8cm]{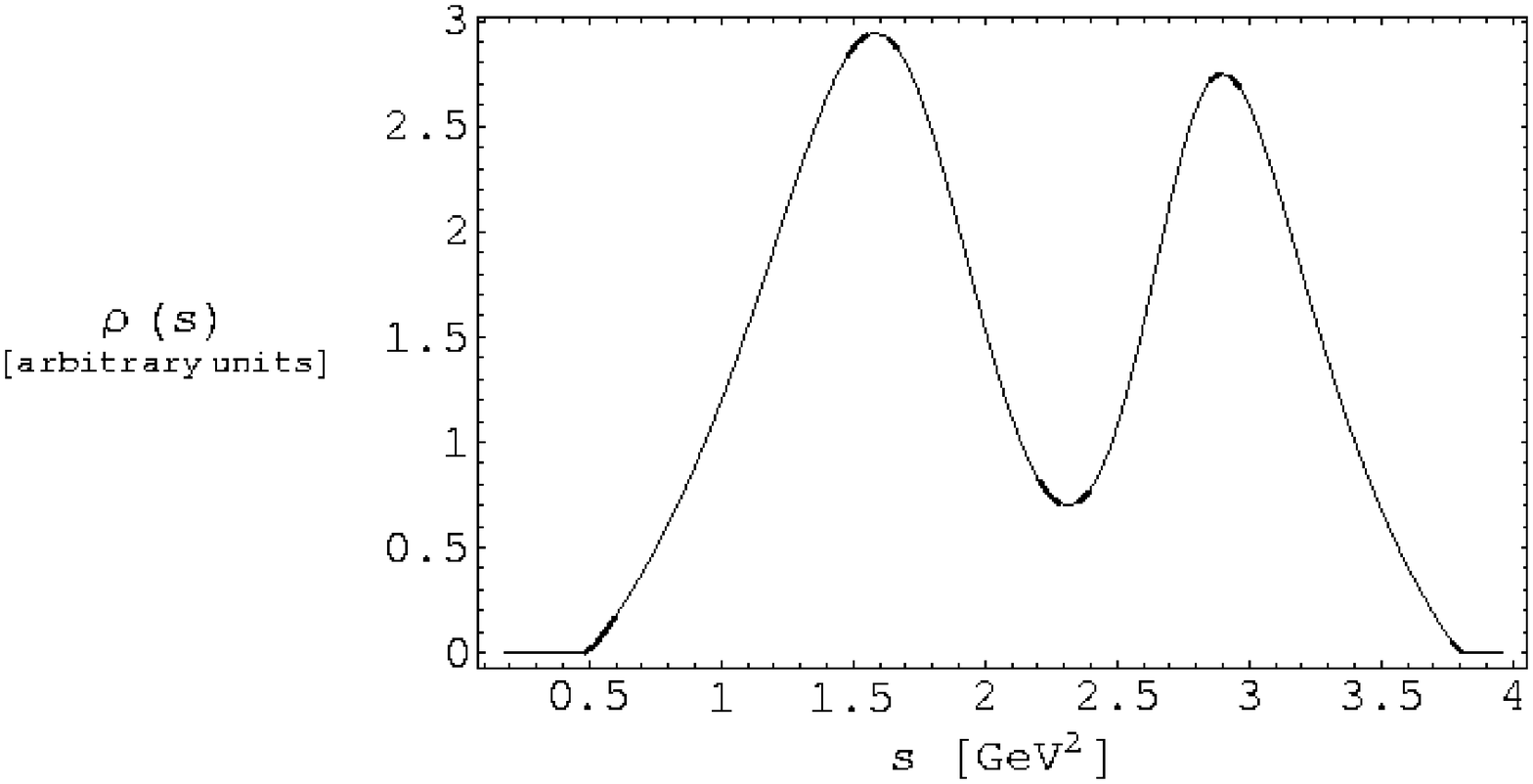}
\end{minipage}\hfill
\begin{minipage}[c]{.30\linewidth}
Fig. 6: A model of the spectral function at intermediate energies.
\end{minipage}
\vspace{0.3cm}
\begin{minipage}[c]{.27\linewidth}
Fig. 7: Duality test for the model exhibited on Figure 6. $R_{Had}$ does not involve
the pion contribution.
\end{minipage}\hfill
\begin{minipage}[c]{.70\linewidth}
\includegraphics*[width=8cm]{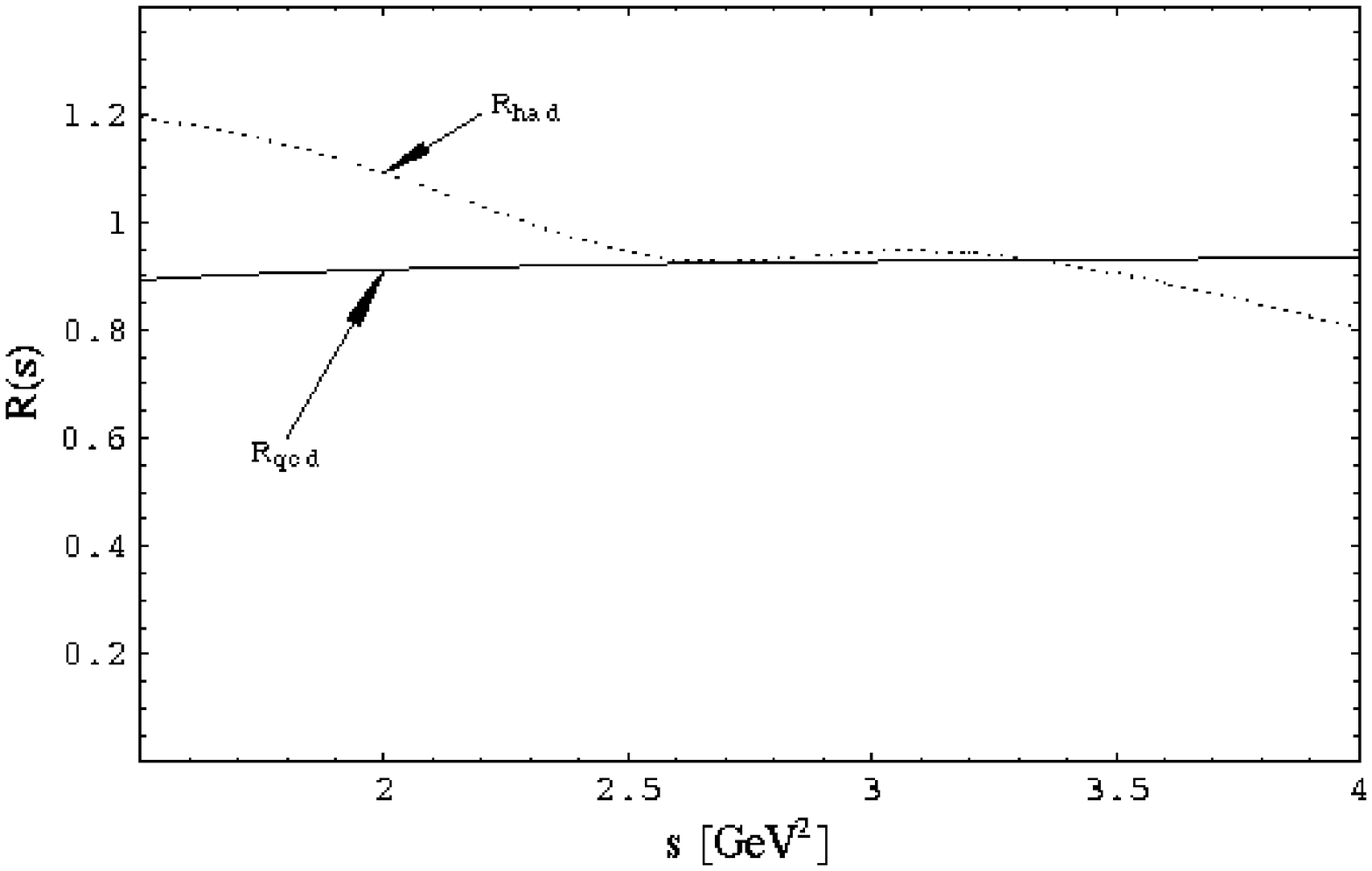}
\end{minipage}
%\end{figure}
\subsection{Relation between quark masses and condensates}
Even if today, QCD sum rules for $\psi^{\bar ud}_{5}$ do not yet allow 
to determine $\widehat m$, they can be combined with the Ward identity
\begin{equation}
	\psi^{\bar ud}_{5}(0)=-(m_{u}+m_{d})<\Omega\vert\bar uu+\bar 
	dd\vert\Omega>\ ,
	\label{psi5}
\end{equation}
and this additional information can be used to eliminate the unknown 
normalization of the spectral function. In this way, one can 
investigate the variation of the condensate \ref{psi5} with $\widehat 
m$, {\bf keeping fixed} \boldmath$F^{2}_{\pi}M^{2}_{\pi}\ $\unboldmath
at its physical value. Notice that \ref{psi5} does not directly 
involve the condensate parameter $B_{0}$ which is defined in the 
chiral limit. For small $\widehat m$, one has, however
\begin{equation}
	\frac{1}{F^{2}_{\pi}}\psi^{\bar ud}_{5}(0)=4\widehat 
	mB_{0}+2\widehat m^{2}C+\ldots\ ,
	\label{1Fpi}
\end{equation}
where $C$ is an ultraviolet counterterm depending on the way 
$\psi_{5}(0)$  is renormalized. One may now consider the Laplace sum 
rule
\begin{equation}
	u\tilde\phi_{5}(u)+\frac{1}{\pi}\int\frac{dx}{x}\exp(-\frac{x}{u})
	\mbox{Im}\ \psi^{\bar ud}_{5}(x)=\psi^{\bar ud}_{5}(0)\ ,
	\label{u5}
\end{equation}
where $\tilde\phi_{5}(u)$ is the Borel-transform of the function 
$\psi_{5}(-Q^{2})/Q^{2}$, together with its u-derivatives. 
$\tilde\phi_{5}(u)$ is assumed to be given by the QCD OPE for $u\geq 
2$ GeV$^{2}$. Using inside these two sum rules a simple {\bf model} 
for the spectral function
\begin{equation}
	\rho(t)=(m_{u}+m_{d})^{2}_{1\ GeV}
	\left\{g^{2}[\delta(t-M^{2}_{1})+\kappa\delta(t-M^{2}_{2})]+
	\theta(t-t_{0})\gamma_{as}(t)\right\}\ ,
	\label{theta}
\end{equation}
where $M_{1}=$ 1300 MeV, $M_{2}=$ 1770 MeV, $\kappa\sim 1$ and 
$\gamma_{as}(t)$ is given by QCD asymptotics, one can eliminate the 
unknown constant $g^{2}$ and for each $\widehat m$ (1 GeV), infer a 
value of the ratio $\psi^{\bar ud}_{5}(0)/2F^{2}_{\pi}M^{2}_{\pi}$. 
In order to control the sum rule stability, individual output values 
of the condensate ratio are displayed in Fig. 8 as a function of the 
Borel-transform variable $u$. Fig. 9 contains a compilation of the 
dependence of the condensate ratio on $\widehat m$ indicating the 
uncertainty arising from the weak dependence on the Borel-transform 
variable. 
\begin{figure}[h]
\begin{minipage}[c]{.70\linewidth}
\includegraphics*[width=8cm]{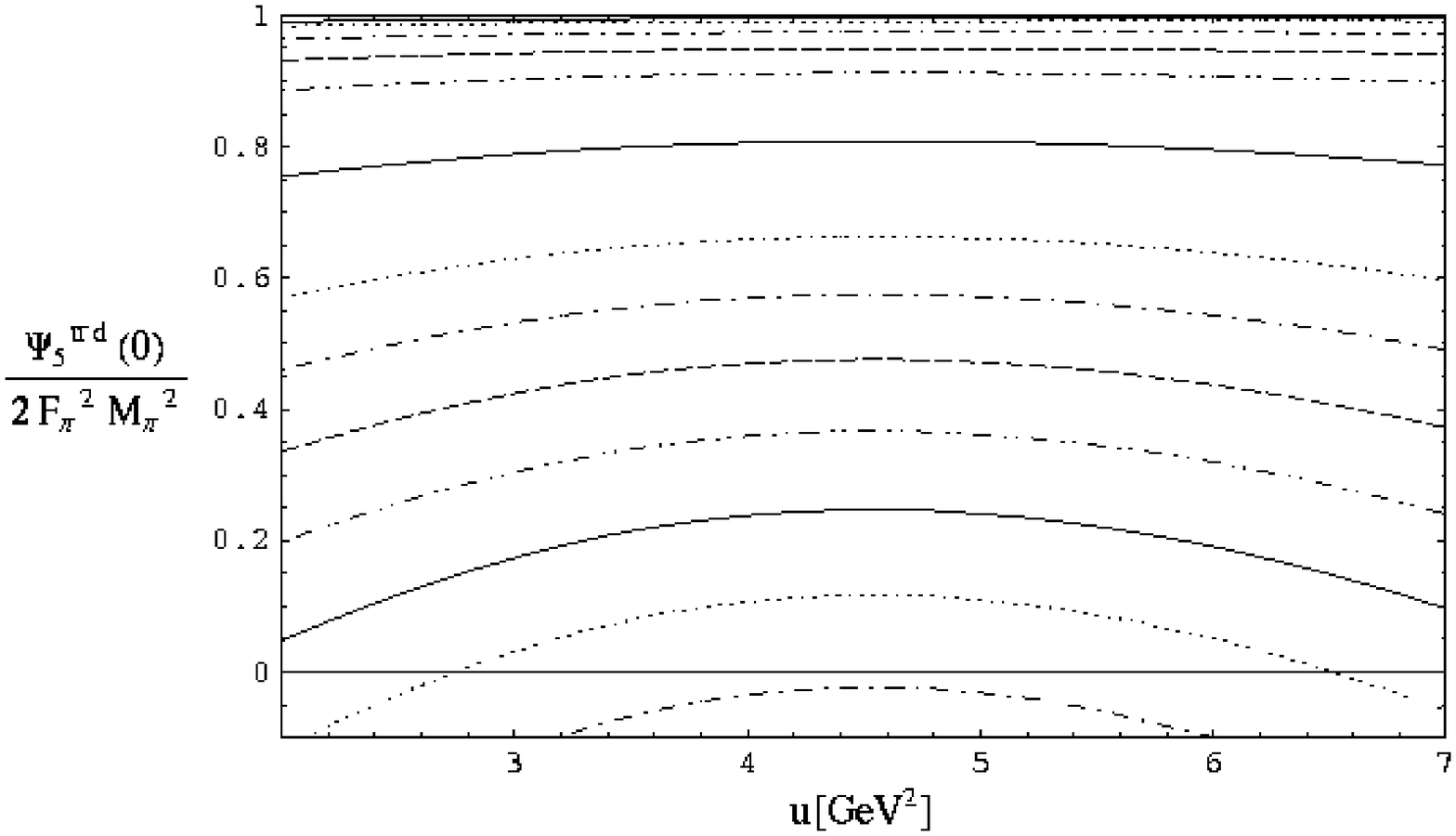}
\end{minipage}\hfill
\begin{minipage}[c]{.27\linewidth}
Fig. 8: Laplace sum rule stability: $u$-dependence of the output condensate values
for different input $\widehat m$.
\end{minipage}
\end{figure}
\begin{figure}[h]
\begin{minipage}[c]{.27\linewidth}
Fig. 9: The condensate ratio as a function of $\widehat m$: $\bar ud$ case.
\end{minipage}
\begin{minipage}[c]{.70\linewidth}
\includegraphics*[width=9cm]{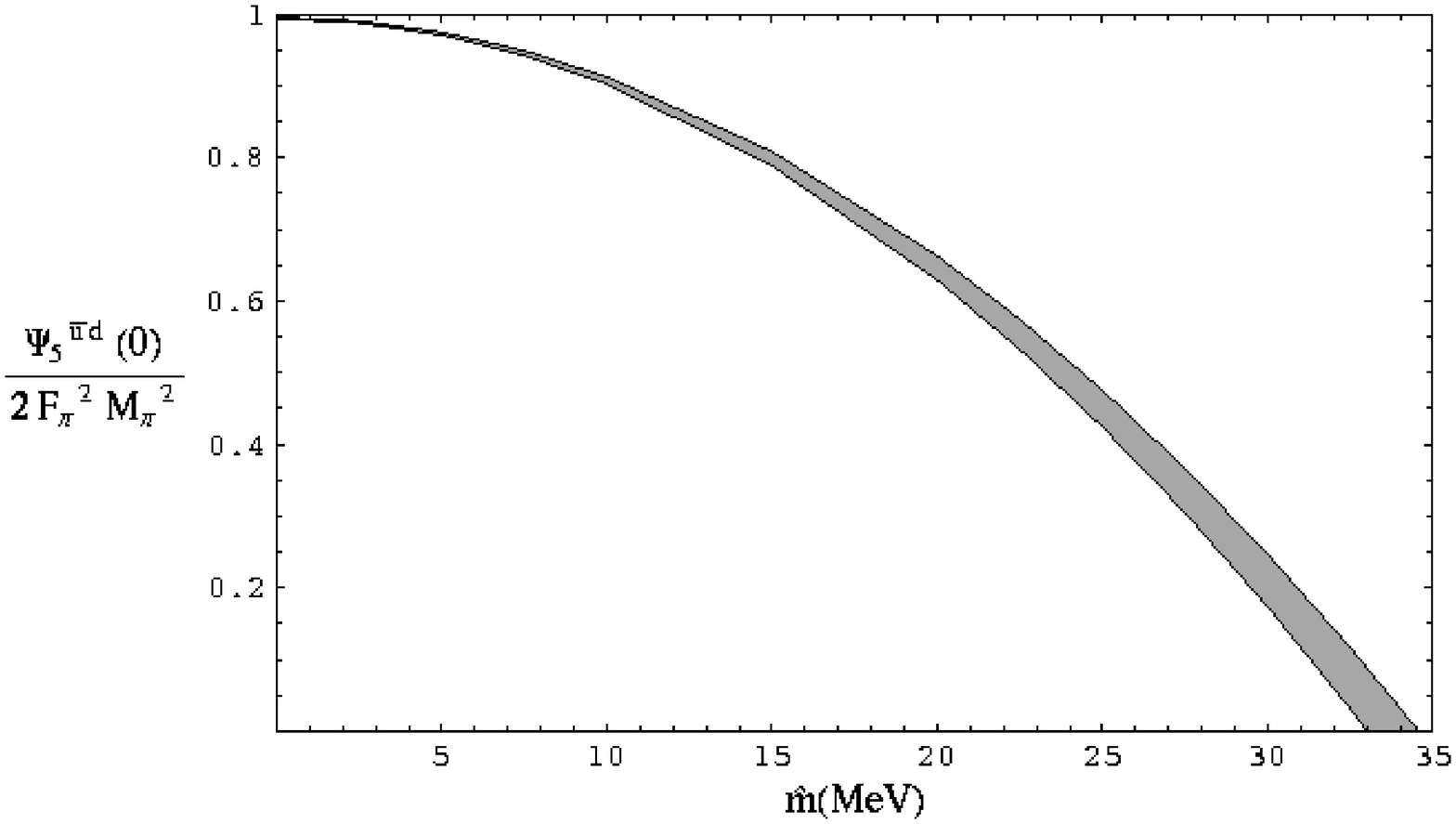}
\end{minipage}
\end{figure}

The shape of this curve can be understood within $G\chi PT$ 
combined with the expansion \ref{1Fpi}. One obtains
 \begin{equation}
 	\frac{\psi^{\bar ud}_{5}(0)}{2F^{2}_{\pi}M^{2}_{\pi}}=
 	1-4\frac{A_{0}-\frac{1}{4}C}{M^{2}_{\pi}}\widehat m^{2}+\ldots\ .
 	\label{1-4}
 \end{equation}
 Comparing with the curve on Fig. 9 one concludes that 
 $A_{0}-\frac{1}{4}C=4.7\pm 0.7$ and remains unaffected if one 
 varies $\widehat m$ and keeps $M^{2}_{\pi}$ fixed. This number should 
 be confronted with the estimates of $A_{0}$ discussed in sect. 4.
 
 The same analysis may now be performed in the $O^{-}$ strange 
 channel: it suffices to replace $\psi^{\bar ud}_{5}$ by 
 $\psi^{\bar us}_{5}$, $\widehat m$ by $\frac{1}{2}(m_{u}+m_{s}),\ 
 F^{2}_{\pi}M^{2}_{\pi}$ by $F^{2}_{K}M^{2}_{K}$ and change the 
 position of resonances in the model \ref{theta} for the spectral 
 function ($M_{1}=1460$ MeV, $M_{2}=1830$ MeV). The stability of the output for 
 various values of $m_{u}+m_{s}$ is shown on Fig. 10 and the final 
 dependence of $\psi^{\bar us}_{5}(0)/2F^{2}_{K}M^{2}_{K}$ on 
 $m_{u}+m_{s}$ is collected in Fig. 11. Notice that 
 $\psi^{\bar us}_{5}(0)=0$ for $m_{u}+m_{s}\simeq$ 250 MeV to be 
 compared with the experimental value of $m_{s}$ \ref{mtau}.

%\begin{minipage}[c]{.70\linewidth}
\begin{center}
\includegraphics*[width=9cm]{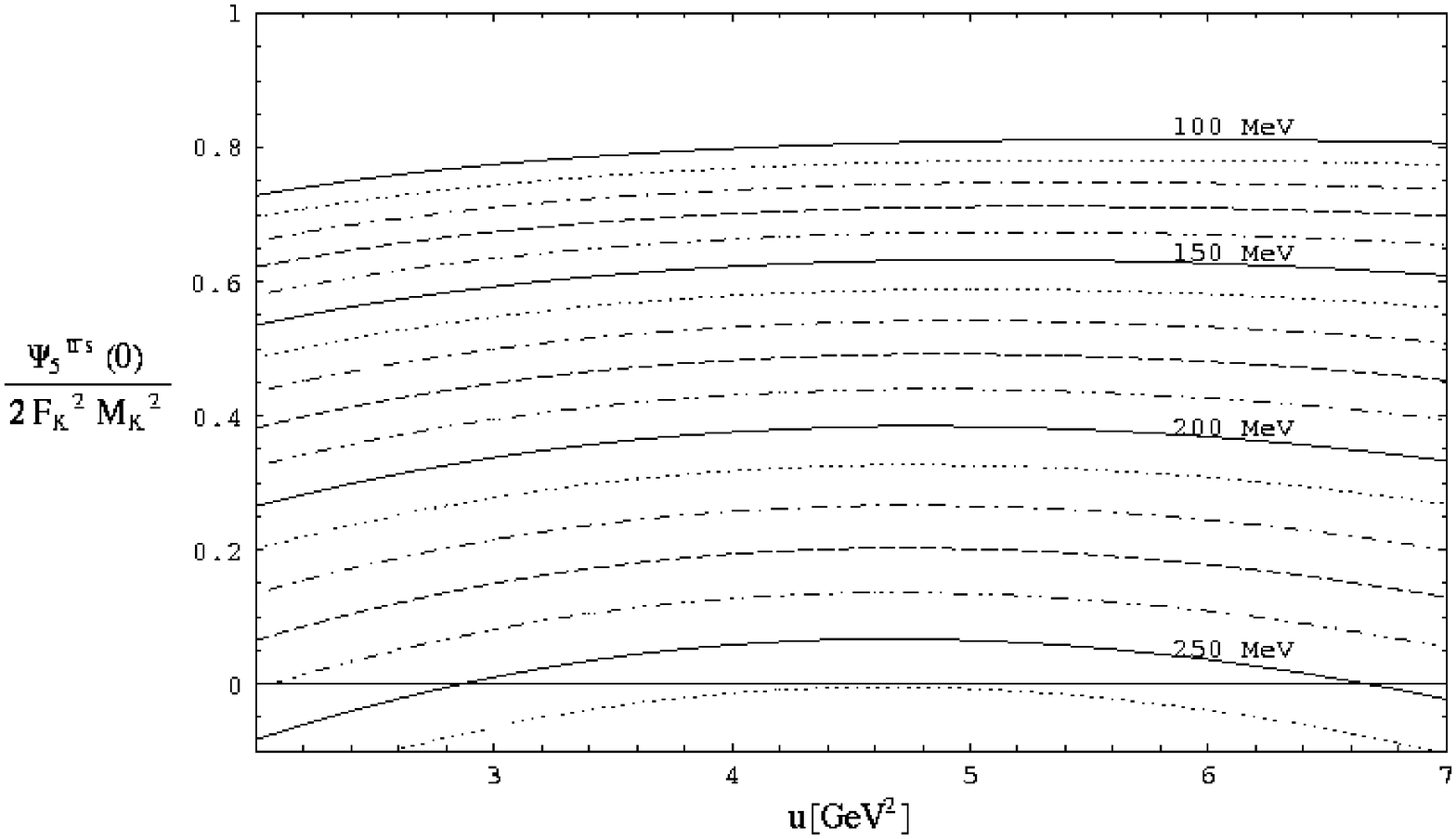}
\end{center}
%\end{minipage}\hfill
%\begin{minipage}[c]{.27\linewidth}
\begin{center}
Fig. 10: Same as Figure 8: the $\bar us$ case.
\end{center}
%\end{minipage}
\begin{figure}[h]
\begin{minipage}[c]{.27\linewidth}
Fig. 11: Same as Figure 9: the $\bar us$ case.
\end{minipage}\hfill
\begin{minipage}[c]{.70\linewidth}
\includegraphics*[width=8cm]{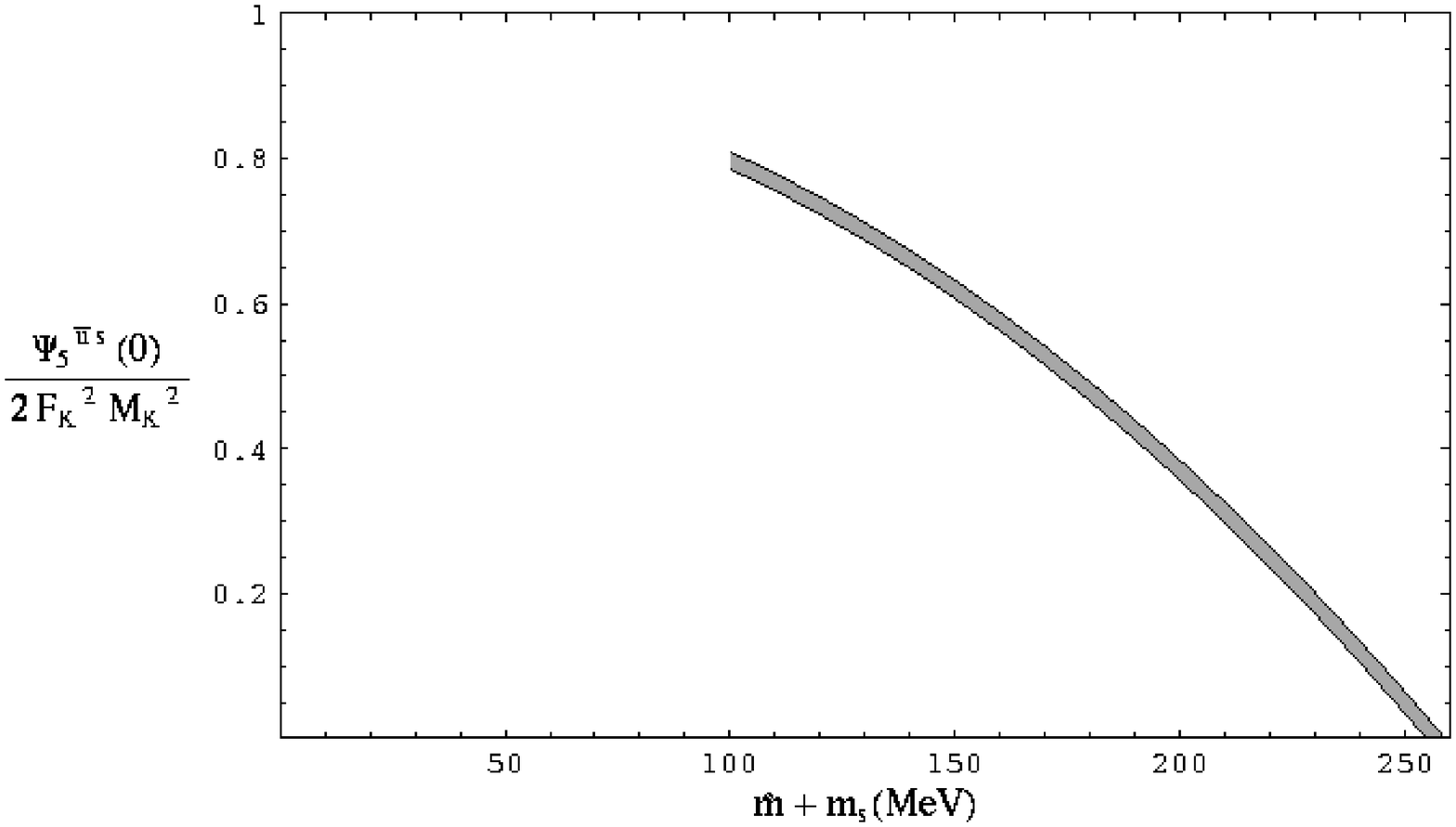}
\end{minipage}
\end{figure}

\section{Experimental tests}
Experimental signatures of the chiral condensate are not easy to 
identify: $<\bar qq>$ enters observable quantities multiplied by 
quark masses and it only manifests itself through tiny symmetry 
breaking effects. $G\chi PT$ provides a theoretical basis for a 
systematic search of possible experimental tests of the importance of 
$q\bar q$-condensation. In this way, the best manner to control 
the GOR ratio $X_{GOR}=2\widehat mB_{0}/M^{2}_{\pi}$ and the 
quark mass ratio $r=m_{s}/\widehat m$ experimentally, has been found 
sofar in the {\bf low-energy} \boldmath$\pi-\pi\ $\unboldmath {\bf 
scattering} \cite{Fuchs91}, \cite{Stern93}. The corresponding scattering 
amplitude at the leading \cite{Weinberg66} and one loop \cite{Gasser83} level 
has to be corrected including the two-loop contribution, in order to achieve 
the theoretical accuracy needed for a decisive test. This has been 
done both in the standard \cite{Bijnens96}, \cite{Bijnens97} and the 
generalized  \cite{Knecht95b}, \cite{Knecht96} frameworks.
\vspace{0.5cm}
\begin{center}
	\includegraphics*[scale=0.4]{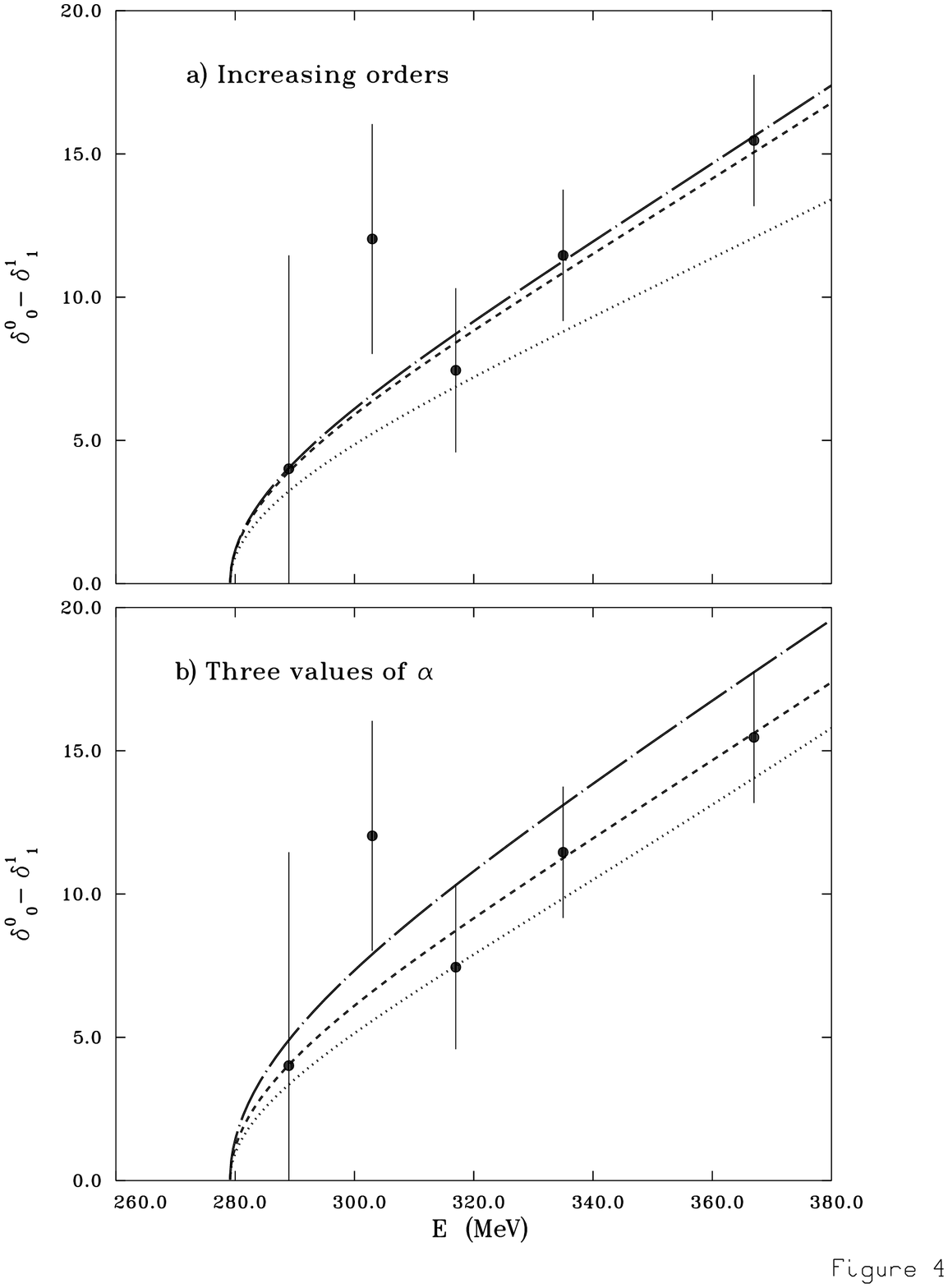}
\end{center}
%\begin{center}
Fig. 12: $\pi-\pi$ phases $\delta^{0}_{0}-\delta^{1}_{1}$ as a function of c.m.
energy. a)~$G\chi PT$ convergence: $O(p^{2}),\ O(p^{4})$ and $O(p^{6})$
orders are compared. b)~$G\chi PT$ predictions: $X_{GOR}\simeq 1$
(dotted), $X_{GOR}=0.6$ (dashed), $X_{GOR}=0.2$ (dash-dotted). Data are from
\cite{Rosselet}.
\vspace{1cm}
The typical 
observable, the phase shift difference 
$\delta^{0}_{0}(E)-\delta^{1}_{1}(E)$ measurable is high statistics 
$K_{l4}$ decays is shown and compared with existing data points 
\cite{Rosselet} on Fig. 12. Fig. 12a illustrates the convergence of 
$G\chi PT$, comparing the tree, one-loop and two-loop results. Fig. 
12b shows the predictions for $X_{GOR}\simeq 1$ (dotted curve), 
$X_{GOR}\simeq 0.2$ (dash-dotted curve) as well as the best fit 
obtained for $X_{GOR}=0.6\pm 0.4$. A need for more precise data 
is obvious. A similar situation is observed with $s$-wave scattering 
lengths: the experimental value $a^{0}_{0}=0.26\pm 0.05$ has to be 
compared with (two-loop) $S\chi PT$ prediction $a^{0}_{0}=0.21$ and a 
value  $a^{0}_{0}=0.27$ corresponding to the case $r=m_{s}/\widehat 
m\simeq 10$. Results of new high precision experiments are awaited: 
more precise $K^{+}_{l4}$ data should come from BNL \cite{Lowe97} and 
from the new $\phi$-factory Da$\phi$ne \cite{Lee97}. 
The experiment "Dirac" at CERN 
aims at a determination of $a^{0}_{0}-a^{2}_{0}$ to 5\%, measuring and 
correctly interpreting the lifetime of $\pi^{+}\pi^{-}$-atoms 
\cite{Schacher97}. It is 
conceivable that in a few years the experimental answer to the 
question of the size of $<\bar qq>$ will be known.

Low energy $\pi\pi$ scattering is not the only possible source of the missing
information: {\bf i)}~The size of the spectral function
$\rho_{3\pi}$, \ref{Im}, can be directly controlled, measuring the 
tiny {\bf azimuthal asymmetries in the decay} \boldmath$\tau\to 
3\pi+\nu_{q}\ $\unboldmath 
\cite{Stern94}. {\bf ii)}~The quark mass ratio $r= m_{s}/\widehat m$ 
can be extracted, comparing the {\bf deviations from the 
Goldberger-Treimann relation} in three different channels 
\cite{Fuchs90}. This test requires an accurate determination of 
strong coupling constants $g_{\pi NN},\ g_{K\Lambda N},\ g_{K\Sigma 
N}$. {\bf iii)}~Additional tests in $K$ and $\eta$-decays are possible.
%\end{center}
\section*{Acknowledgements}
I am indebted to Norman Fuchs, Marc Knecht and Bachir Moussallam for a
collaboration during a preparation of this talk. Discussions with Juerg
Gasser, Heiri Leutwyler and Andrei Smilga have been extremely helpfull. Aron
Bernstein, Dieter Drechsel and Thomas Walcher should be acknowledged for a
perfect organization of the workshop.
\bibliographystyle{alpha}

\newcommand{\etalchar}[1]{$^{#1}$}

\end{document}